\documentclass[preprint,12pt]{elsarticle}
\usepackage{lineno,hyperref}
\usepackage{float}
\usepackage{placeins}
\modulolinenumbers[5]
\journal{arXiv}

\begin{document}

\begin{frontmatter}

\title{Blockchain Education: Current State, Limitations, Career Scope, Challenges, and Future Directions}

\author[rvt]{Rizwan Patan\corref{cor1}}
\ead{prizwan5@gmail.com}
\author[rvt]{Reza M. Parizi}
\ead{rparizi1@kennesaw.edu}
\author[focal]{Mohsen Dorodchi}
\ead{Mohsen.Dorodchi@uncc.edu}
\author[rvt]{Seyedamin Pouriyeh}
\ead{spouriye@kennesaw.edu}
\author[focal]{Audrey Rorrer}
\ead{Audrey.Rorrer@uncc.edu}

\address[rvt]{Decentralized Science Lab, Kennesaw State University, GA, USA}
\address[focal]{University of North Carolina at Charlotte, NC, USA}
\cortext[cor1]{Corresponding author}

\begin{abstract}
Blockchain is a revolutionary technology, and its growth started in various industries (such as IT, education, business, banking, and many others) to capitalize on it. Currently, in higher education institutions (HEIs) adoption of blockchain education needs to be improved in the academic programs and curriculums. In addition,  HEIs must make many intense changes in the teaching and learning methods to educate learners about blockchain technology and its applications to meet the current industry workforce demand. Due to a lack of academic programs and courses, students nowadays rely on online resources and pay non-academic organizations a high fee. This paper provides a comprehensive survey of blockchain education's current state of the art by reviewing the different academic programs and industry workforce demand. In addition, blockchain application trends which includes the market growth and demands are discussed. Moreover, the blockchain career scope for different disciplines of students is examined. 

 
\end{abstract}

\end{frontmatter}
\section{Introduction}
\label{sec:intro}
Blockchain technology is one of the emerging areas nowadays that has gained widespread attention over recent years. Much revolutionary growth started in various industries (such as IT, education, business, banking, and many others) to capitalize on this technology. Blockchain act as a decentralized database, including transaction data that is permanently saved and encrypted. The governing architecture of a blockchain is a node, a laptop, a computer, or a server that has a complete copy of the blockchain's transaction history rather than depending on a central authority to store and validate data. A peer-to-peer network of nodes forms a blockchain to constantly share the most recent data block to keep all nodes in sync \cite{R51}. Due to these characteristics, many blockchain-based applications are being developed (from Defi \cite{R1}, NFTs \cite{R2}, Dapps \cite{R3}, and supply chains \cite{R4} to rewards and more) to disrupt traditional business models and create unique interest. Blockchain technology's recent rapid expansion forced many industries to redesign and rethink many fundamental components of existing systems. Blockchain is setting a new trend by innovating businesses by delivering robust characteristics such as trust, security, privacy, and identity \cite{N1}. The best way to launch a blockchain application requires technical expertise and experience to maximize the chances of a successful, thriving implementation, and it is only possible with strategic design \cite{R5}.

The blockchain academy provides the following reasons why embracing blockchain-specific education in higher education institutions (HEIs) is important \cite{R13}: \textit{"We are noticing an increasing demand of current workforce that firmly pushes blockchain education in HEIs. In order to introduce new programs and individual courses is essential to foster skills to meet industry standards."} Nowadays, various industries are redesigning their existing system and launching new startup businesses with blockchain. 

Blockchain technology has become one of the most competitive spaces around the globe \cite{R45}. Unlike other technologies, blockchain technology has adapted to different applications. Nurturing students in blockchain education \cite {R8} is an ongoing process since technology is evolving rapidly. However, HEIs must keep their curriculum up to date, continuously develop students' skill sets, and provide real-time hands-on experience; otherwise, their skills may become obsolete. HEIs can find several suitable practices that can build to improve students' proficiency in specific tech niches. These credentials can portray a better image of their students in the visions of an employer.

Blockchain education survey conducted by CoinDesk \cite{R14} in 2021 to rate the top 20 universities in the world that offer the best blockchain education. In this top list, out of 20, 14 schools are private with limited courses. Between 2018 - 2021, MIT, Harvard, and many other reputable institutions only had the highest student placements and publications in the blockchain area. Even long while, the blockchain communities strive to promote decentralization, and open access projects make available for students to advance this technology. This widespread adoption of its education establishes a hierarchy between academia and industry. Industries using blockchain technology require more than 300 \% of the existing workforce, so HEIs may need to develop and adopt blockchain courses that will take two to three years to fulfill current workforce demand \cite{R12}.
\subsection{Motivation}
As much as blockchain technology has advanced over the years, universities still need to catch up to implement comprehensive courses that can keep up with the evolving industry demand \cite{R7}. Offering courses often need to reflect emerging changes in pedagogy and therefore lack real-world business context \cite{R7}. According to the BlockGeeks (one of the largest blockchain education hubs on the internet), 2022 survey reports \cite{R8}, while the demand is growing for blockchain experts (who help to infuse blockchain in building and designing solutions that leverage cross-asset frameworks and concepts), there is a lack of curricular educational resources to educate undergraduate and graduate STEM students.
\subsection{Goal of the Study}
The main goal of this study is to create awareness and establish the need for new programs with novel pedagogy for blockchain education.  The current consensus is that blockchain education curricula in HEIs are mainly limited \cite{R19}. This is because they mainly teach and train very high-level materials around the hype and motives for adopting blockchains or very low-level smart contracts programming skills without emphasizing blockchain system design . In addition, flagship schools have developed very specialized courses due to the growing needs that could be more easily adaptable to the large undergraduate and graduate population \cite{R9}. However, as a common pattern to a fast-growing technology, such effort in HEIs requires curricular innovation that is strategic, scalable, and sustainable. A proper pedagogical approach to blockchain education based on "abstraction" would provide the necessary bridge between the technology and business levels \cite{R76}.
\subsection{Contributions}
Our main intention with this paper is to show the current state of blockchain education, industry growth, and workforce demand. The following lists the key contributions:
\begin{enumerate}
\item We investigate various research-led and industry-led education standards that are best suitable to educate blockchain in HEIs.
\item We investigate blockchain market growth in recent years and the type of workforce demanded by various industries. 
\item We provide necessary directions for HEIs to innovate their programs and courses to meet the current industry workforce demand.
\end{enumerate}

The rest of the paper is divided into five different sections as follows. Section \ref{sec:BS} background study discusses the current state of blockchain education. Section \ref{sec:CS} shows the career scope for different streams of blockchain learners. Section \ref{sec:C} provides the blockchain application trends which includes industry growth and workforce demand. Section \ref{sec:FD} gives future directions of blockchain education and its scope based on blockchain profession. Finally, Section \ref{sec:con} is the conclusion that summarizes the overall paper.

\section{Background Study}
\label{sec:BS}
According to the 2021 report by the Office of Educational Technology in the United States \cite{R10}: \textit{"Currently considering a growing demand of the workforce and startup needs are stating that blockchain education and its skills will play an important role shortly and evolving to lead industries."} As per this report, blockchain adoption within HEIs is highly significant. Moreover, providing sufficiently structured teaching and training models and resources within higher education is essential to support the infusion and expansion of blockchain technology in computing and engineering and a well-structured connection to business, finance, and related areas. Moreover, HEIs require a new-age blockchain curriculum to support and guide students and produce a skilled workforce along with entrepreneurs in the blockchain area.

In order to train students to become blockchain professionals, it is highly vital to have a strong understanding of the fundamental concepts and terminology used in the blockchain industry \cite{R6}. One cannot advance in the blockchain sector without possessing this talent, as it is a prerequisite, and without having adequate subject expertise. Therefore, it is important to educate students on the fundamental ideas behind blockchain technology, develop a more in-depth and distinct comprehension of these ideas, and educate students on real-time blockchain applications and the mechanisms underlying various consensus protocols. Investigate the most recent developments that have been made using blockchain technology \cite{R17}. Taking classes from a training provider officially recognized as fulfilling certain standards is the best way to acquire useful and marketable skills. In this regard, we discovered that the lessons and training programs made available resources like Simplilearn, blockchain council, and many others are straightforward to comprehend \cite{R24}. In addition, they provide a training program covering many different topics and the fundamental technical core that needs to become a master in this area.

Blockchain professionals such as developers and architects are interested in understanding the widespread frenzy surrounding blockchain. Bitcoin and cryptocurrencies are the target audience for the blockchain certification training that Simplilearn has developed. Students will become familiar with the fundamental architecture and underlying technical mechanisms of blockchain platforms such as Bitcoin, Ethereum, Hyperledger, and Multichain. They will also use the most up-to-date tools to construct applications, establish their private blockchain, deploy smart contracts on Ethereum, and gain hands-on experience with projects \cite{R57}. Students will receive a certification in blockchain development once students have finished the course. Students will then be prepared to take on the challenges presented by this fascinating new technology. Simplilearn is prepared to be a helpful resource for students not only to become a blockchain developer but also to provide additional training and skills in topics related to blockchain development, such as DevOps, Software Development, and Cloud Computing. Simplilearn stands ready to be a valuable resource for students. Suppose students recall the earlier discussion on ethical hackers. In that case, students will be pleased to find that Simplilearn even provides a training course in Certified Ethical Hacker v10 for those interested in obtaining this certification \cite{R26}. Explore the Simplilearn platform to embark on a profession that is both exciting and lucrative. The intention of providing this information is to show how far away the external nonacademic organizations are considering providing value for blockchain education. However, HEIs must catch up to meet this demand by fetching novel pedagogy approaches \cite{R20}. 

Table \ref{c2:tab1} summarizes our findings of the current list of universities in the United States that offer blockchain-related courses \cite{R27}. By analyzing the types of courses offered by these institutions and demonstrating the dearth of blockchain education, it seems that only courses 3 and 20 have more aspects of architecture and development. In particular, the courses offered by the institutions (courses 1 to 3) only cover the general concepts of blockchain architecture and mostly focus on the most common business applications such as smart contracts and cryptocurrency and lack a structured approach in line with business needs from a development and architect standpoint. Flagship schools have developed specialized courses due to the growing needs, but they are not easily adaptable to the large undergraduate and graduate population.

\begin{table}
\tiny
\centering
\caption{Some Universities in the US that Offer Blockchain-related Courses}
\label{c2:tab1}
\resizebox{\textwidth}{!}
{
\begin{tabular}{|p{1.8in}|p{1.5in}|p{1.5in}|}
\hline
\multicolumn{1}{|c|}{\textbf{Course/certificate}} &
  \multicolumn{1}{c|}{\textbf{University}} &
  \multicolumn{1}{c|}{\textbf{Offered By}} \\ \hline
1. Fundamentals of Blockchain and Smart Contracts
\newline 2. Blockchain for Business &
Kennesaw State University &
Computing and Software Eng. 
College of Business \\ \hline
3. Blockchain System Architecture &
UNC, Charlotte &
Department of CS \\ \hline
4. Courses in Entrepreneurial Finance \newline
5. FinTech Law and Policy \newline
6. Blockchain Business Models \newline
7. Certification for Blockchain Applications &
Duke University &
Department of Engineering, CS, math, statistics \\ \hline
8. Introduction to Blockchain and Bitcoin \newline
9. Breakthrough Innovation with Blockchain Tech.\newline
10. Certification in FinTech &
Harvard University &
Department of CS \\ \hline
11. Blockchain and Industry \newline
12. Blockchain Foundations and Frameworks \newline
13. Blockchain Cases &
Loyola Marymount University &
Department of CSIT \\ \hline
14. Blockchain Ethics \newline
15. Blockchain Technologies &
MIT &
MIT Management Sloan School \\ \hline
16. Blockchain for Business &
Portland State University &
School of Business \\ \hline
17. Bitcoin and Cryptocurrency Technologies &
Princeton University &
Department of CS \\ \hline
18. Blockchain Technologies for Business Leaders &
Rutgers University &
School of Law \\ \hline
19. Blockchain and Cryptocurrency &
Stanford University &
Department of CS \\ \hline
20. Decentralized Applications, Smart Contracts, Blockchain Platforms, and Blockchain Basics &
SUNY at Buffalo &
Department of CSE \\ \hline
21.BT and Bitcoin and Cryptocurrencies \newline
22.Certification in Blockchain Fundamentals &
UC, Berkeley &
School of Business \\ \hline
23. Blockchain, The Merkle Tree and Cryptocurrencies \newline
24. The Blockchain System, Cryptography  Hashing Overview and specialization  &
University of California, Irvine &
Division of Continuing Education \\ \hline
25. Cryptocurrency and Blockchain; specialization course in Foundations and Applications &
University of Pennsylvania &
Business, law, and finance departments \\ \hline
\end{tabular}
}
\end{table}

Blockchain is presented as the next “big” digital technology trend that will challenge HEIs to create new courses and programs \cite{R27}. In contrast, HEIs are currently slow to adopt blockchain education in their academic programs compared with industry demands \cite{R29,R30}. Since the technology is relatively new and complex, the infusion of blockchain into the existing undergraduate curriculum is challenging. Though some independent courses exist, there is a real need to develop new curriculums and innovative teaching and learning approaches that are modern, adaptable, scalable, and adaptive to industry needs.

Currently, the U.S. Department of Education Blockchain Action Network and the American Council on Education’s Education Blockchain Initiative (EBI) \cite{R22} are actively exploring the teaching and training of blockchain technology and how it can potentially revolutionize the field \cite{R31,R32}. The Education Blockchain Action Network brings educators, administrators, students, and technology developers to collaborate in examining how institutions could shape the future of blockchain education \cite{R33,R34}. In addition, the network serves as a hub for stakeholders in the blockchain sphere as a “shared, community-driven, action-oriented space” for discussion and open-source project development. However, existing blockchain curriculum's in HEIs accommodate and teach only a few beginner-level courses \cite{R35,R36}. Few institutions are now offering advanced courses in blockchain, but this count is minimal compared to today’s blockchain industry demand.

The consensus, however, is that HEIs need to customize their programs and courses based on blockchain enterprise needs for different career tracks \cite{R38}. To keep up with this rapidly evolving field, in order to create the trend, HEIs need to adopt new teaching and learning strategies. Update significant portions of their blockchain syllabi every semester, which would demand multiple learning modules beyond a semester or two to fulfill the current industry workforce demand \cite{R39}. Moreover, most HEIs focus on basic-level studies in the blockchain area and do not pay attention to high-demand fields like developers, engineers, system designers, and solution architects \cite{R40}. Moreover, several non-academic organizations such as \textit{Blockchain Council, Udemy}, and many others have started offering short-term training courses with limited activities. Details of a few reputed courses or certifications with their characteristics are shown in Table \ref{c2:tab2}.

\begin{table}
\centering
{\caption{\centering Analysis of several non-academic organizations offering development and architect courses} }
\label{c2:tab2}
\resizebox{\textwidth}{!}
{
\begin{tabular}{|c|p{0.4\linewidth}|c|c|c|c|c|}
\hline
\multicolumn{1}{|c|}{\textbf{No}} &
  \multicolumn{1}{c|}{\textbf{Course/Certification Title}} &
  \multicolumn{1}{c|}{\textbf{Organization}} &
  \multicolumn{1}{c|}{\textbf{Training Level}} &
  \multicolumn{1}{c|}{\textbf{Practical Env.}} &
  \multicolumn{1}{c|}{\textbf{Community}} &
  \multicolumn{1}{c|}{\textbf{Fee (\$)}}\\ \hline
1& Blockchain Development on Hyperledger Framework & Udemy & Beginner & NA & NA & 16.99 \\ 
\hline
2& 
Certified Blockchain Security Expert (CBSE) & 101 Blockchains & Intermediate & NA & NA & 399 \\ 
\hline
3& Blockchain Specialization & SUNY & Advanced & NA & NA & 1200 \\ \hline
4 & Developing Blockchain Applications & RMIT University & Intermediate & NA & NA & 993 \\ 
\hline
5& Master of Blockchain-Enabled Business& RMIT University & Advanced & NA & NA & 23386 \\ 
\hline
6 & Certified Blockchain Architect & Crypto Research Lab & Intermediate & Hands-on & NA & 1695 \\ 
\hline
7& Certified Blockchain Architect & Blockchain Academy & Intermediate & Hans-on & NA & 1800 \\ 
\hline
8& Blockchain Technology and Architecture Lab & EZY skills & Beginner & NA & NA & 199 \\ 
\hline
9& Certified Blockchain Architect & Blockchain Council & Intermediate & NA & Yes & 299 \\ 
\hline
10& Certified Blockchain Solution Architect & Blockchain skill Alliance & Intermediate & NA & NA & 499 \\ 
\hline
11& Certified Enterprise Blockchain Architect & 101 blockchains & Intermediate & Hands-on & Yes & 1200 \\ 
\hline
12& Blockchain for Solution Architect & Knowledgehut,  upGrad & Intermediate & Hands-on & NA & 950 \\ 
\hline
13& Certified Blockchain Solution Architect & Stone Revier eLearning & Intermediate & NA & NA & 1200 \\ 
\hline
14& Blockchain Solution Architecture Training & NTUC LearningHub & Intermediate & Hands-on & NA & 850 \\ 
\hline
15& Enterprise Blockchain Architect Course & Udemy & Beginner & NA & NA & 100 \\ \hline
\end{tabular}
}
\end{table}

Despite the growing importance of blockchain in academia and industries, there is currently no well-developed educational strategy to prepare the current workforce and future students for careers in this field \cite{R11}. Many initiatives are underway to meet this demand, including massive open online courses (MOOCs) (like Coursera's), online courses (like blockchain council), and degree/certificate programs (like those listed in Table \ref{c2:tab1}). Despite the widespread availability of these resources designed for students of varying educational and professional pursuits, practitioner demand remains high.

{\raggedright We pinpoint three main causes for the shortfall in blockchain education:}
\begin{enumerate}
    \item While MOOCs allow students to learn at their own pace, many people drop out because they do not have the proper resources to practice pursuing their education outside the workplace.
    \item The time and effort required to complete established courses/certifications take too much time for working professionals. In addition, difficult for those who lack of blockchain fundamental knowledge.
    \item There are many students are taking a back step and thinking multiple times about a blockchain career by considering and examining the difficulties former students face.
\end{enumerate}

{\raggedright {\textbf{Tips to close this growing gap}}} 

Since students come from various educational backgrounds, no "one-size-fits-all" approach can be taken while designing blockchain learning programs. Micro courses, tutorials, workshops, hackathons, boot camps, and internships are some alternative and flexible modalities we should consider adopting in addition to traditional, classroom-based instruction to meet the needs of students from various backgrounds. These can be used individually or in tandem to create various educational programs. For example, workforce development can benefit from micro courses, workshops, and tutorials; advocacy can use hackathons; internship possibilities enhance degree programs and can be included in the curriculum to improve learning and give practical experience. Creating and distributing blockchain-specific degree programs for college students is the most important factor in spreading technology awareness.

However, for graduate-level programs, many reputed HEIs are offering blockchain worldwide. Such programs will potentially change  businesses in the modern world, including education. However, for graduate-level programs, many reputed HEIs are offering blockchain worldwide. Such programs will potentially change businesses in the modern world, including education. In recent years HEIs in the United States (US) and other countries that started offering graduate programs in blockchain areas have increased, and more students are showing interest by seeing its value. Reputed blockchain universities in the US educate students on the practical applications that serve as the foundation for a successful career in business. 

\section{Career Scope}
\label{sec:CS}

Blockchain technology demand is rapidly growing in various industries, including IT, banking, education, business, and many others. LinkedIn, Glassdoor, Gartner research, and many more survey reports \cite{R36, R39, R43} indicate around a 300\% increase in blockchain-related jobs in the US during 2021. Since then, this demand graph has shown positive growth in the industry \cite{R42}. In addition, a LinkedIn survey report disclosed \cite{R36} those job posts featuring terms like bitcoin, blockchain, and other digital asset-related functions increased 395\% in 2021 in the United States compared to the previous year. This had the most rapid growth in the technology industry, exceeding the whole sector by 4x with a 98\% increase. The biggest number of job ads in the blockchain domain on the employment-oriented social networking service were in banking and software, the two industries that have seen the most growth in blockchain positions over the last five years.

This is a golden time to be a blockchain developer, as prospects in the blockchain area are expanding at an unprecedented rate. The scope of blockchain in the United States and other countries is significantly wider than ever, and it is expected to grow even more shortly \cite{N8}. Following the bitcoin bull market of 2017, the world's top and notable firms are searching for skilled blockchain developers. In 2022, the demand for blockchain engineers increased by more than 517\% over the previous year, 2021. According to the International Data Corporation (IDC), global spending on blockchain technologies will reach \$11.7 billion by 2022. According to IDC, blockchain will develop fastest between 2017 and 2022 \cite{N9}. The breadth of blockchain in the United States is expanding as many organizations look for new ways to use blockchain technology to advance their businesses.

\begin{figure}[htp]
  \centering
  \includegraphics[width=5.2in]{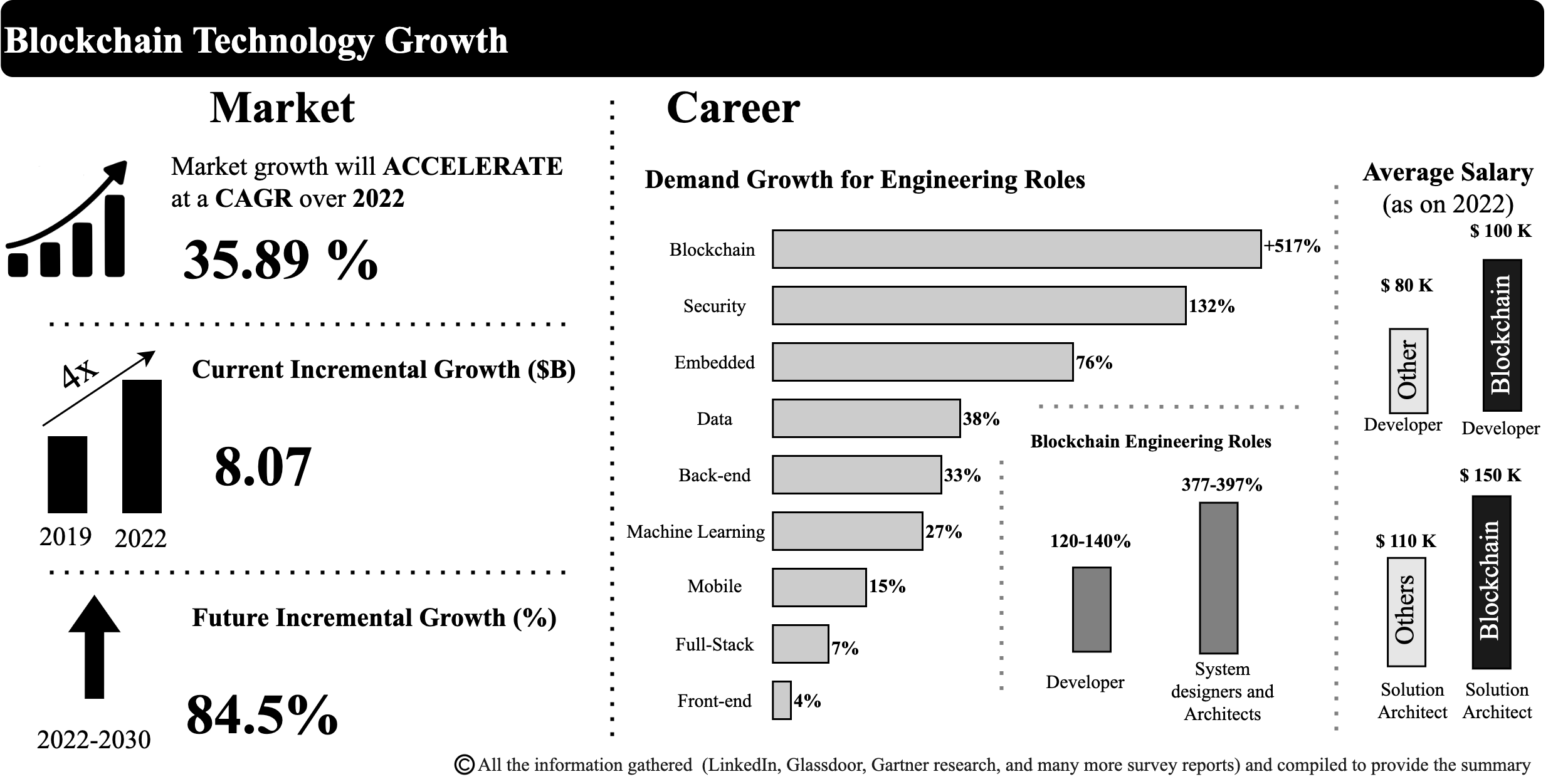}
  \caption{Summary of blockchain industry growth and demand}
  \label{c1:f1}
\end{figure}

Figure \ref{c1:f1} shows a detailed summary of blockchain industry growth and demand in terms of market scale and career goals. In 2021, many industries adopted blockchain technology, including IT, banking, and others. This information is compiled from multiple sources, including LinkedIn, Glassdoor, Gartner research, and many more survey reports \cite{R36, R39, R42, R43, R36}. There are various career tracks for students who plan to study blockchain technology, and recommendations are listed below based on the career track. 

\subsection{To become blockchain developer}

 Blockchain developers are professionals who work with blockchain technology and are in charge of activities like inventing blockchain protocols, writing smart contracts, and many more \cite{R48}.

There are two sorts of blockchain developers: Blockchain Software Developers and Core Blockchain Developers. The following are responsibilities for each profile:
\begin{enumerate}
\item Core Blockchain Developers
\begin{itemize}
    \item Create Blockchain protocols along with consensus protocols.
    \item Network architecture design and monitoring.
    \item Implement blockchain features and functionality, and many others.
\end{itemize}
\item Blockchain Software Developers
\begin{itemize}
    \item Creates blockchain integration APIs.
    \item Develop Decentralized Applications' frontend and backend.
    \item Develop and implement smart contracts and many others.
\end{itemize}
\end{enumerate}

Since blockchain technology is continuously evolving, it is challenging to cope with this rapidly evolving technology as a developer. For example, this paper's first author created Solidity smart contracts for training purposes. These contracts were built using the truffle 4.x toolchain. He came back after a few months to install the same truffle on another Laptop to start fresh. By default, the npm install recommended truffle the latest version to install, i.e., version 5.x. He noticed none of his old code was working, and there were drastic changes from 4.x to 5.x. He had to figure out, understand and update these changes and migrate all the code to 5.x version to make it work. Imagine how difficult it would have been to follow up with these changes for a new learner. This is just an example of Ethereum Smart contracts and Solidity, but the other things also keep changing; hence, as experienced developers and professors, students should always keep a tab on what is happening around the field. This is not fun; most of the time, we need to spend this much time figuring out how these updates impact existing stuff.

However, fewer articles, videos, and presentations talk about the business aspects of a blockchain and how it will impact the industry. There needs to be more information about the core technical aspects of a blockchain, how it works, and more \cite{R67}. Even though there are many resources, they could be high-level or abstract. Although innumerable platforms are available online to learn blockchain, it is very important to pay attention to the quality of knowledge students are gaining. Most institutes claim to offer top-notch courses and certifications for learning blockchain. However, when delivering knowledge of blockchain, students often need help to succeed. HEIs must pay attention to preparing students for different blockchain roles.


As blockchain is a new technology, there are very few developers for blockchain development. Only 0.80\% of the current developers know blockchain development. Due to this, industries pay higher rates for blockchain developers. This rate varies from region to region and country to country. Students can find blockchain developers in many major countries like the USA, UK, Australia, Germany, and others. The average hourly rates will be between \$80- \$150; however, this also differs in many countries.

\subsection{To become blockchain (solution) architect}
Blockchain architects are hot in demand and low in availability. As a result, recruiters are looking for blockchain architects, and they are willing to pay higher salaries than other IT experts.

A blockchain architect provides end-to-end solutions to businesses based on blockchain technology and contributes to developing an entire blockchain ecosystem engagement strategy \cite{R50}. Organizations are looking for specialist individuals that can able to perform below tasks as follows (but not limited to):
\begin{enumerate}
\item Identify, assess, and solve complex business problems.
\item Understand how blockchain architecture, private keys, and standards/ protocols work.
\item In-depth understanding of business components and technical elements of blockchain architecture
\item Responsible for building or integrating with the business logic application software and architecting complex solutions.
\item Know which blockchain platform/protocol provides the best performance and scalability for a given use case.
\end{enumerate}

In terms of salary package, Glassdoor reports that the average base pay for a \textit{Blockchain Architect} average salary in the United States is about \$118,000 per year, and in India is about INR 11,50,000 per year \cite{N11}. Therefore, students with experience in blockchain development want to become successful blockchain architects. We recommend students to Blockchain Council \cite{N12}. which offers online training and certifications in the blockchain space.

The blockchain architect certification courses  listed in Table \ref{c2:tab2} from 6 to 15 will provide students with competitive knowledge of blockchain architectures. These certifications will help students become master the concepts of blockchain architecture tools, business components, and technical elements of blockchain architecture, how to translate requirements into functions, architecting their blockchain solutions, and more \cite{R63}. In short, it will help students acquire skills to pursue their careers. However, these certifications are limited to teaching and training already experienced persons, not providing broad knowledge to basic learners.


Businesses across industries ranging from finance to manufacturing to healthcare know blockchain technologies are evolving. Furthermore, they know the emerging tech has the potential to provide numerous business benefits, such as lower operational costs \cite{R68}, faster transaction speeds, and the elimination of the cost and necessity of using intermediaries to facilitate financial transactions. Besides developers and architects, there are many exciting career options for those skilled in blockchain. There are other prominent roles as well for blockchain professionals are listed below. 
\begin{enumerate}
\item Blockchain Intern
\item Blockchain Engineer
\item Blockchain Designer
\item Blockchain Project Manager
\item Blockchain Quality Engineer
\item Blockchain Legal Consultant or Attorney
\end{enumerate}

\section{Blockchain Application Trends}
\label{sec:C}
We discussed the blockchain market growth and key areas in various sectors for analyzing blockchain workforce demand to discover the key areas of blockchain education that need to focus on more. However, various universities worldwide are providing education on blockchain technology. For example, the blockchain institute of technology (BIT) is one of the world’s leading training and education providers in blockchain technology and cryptocurrency.

These institutions also take many initiatives to offer online blockchain training, but only some colleges and universities offer hands-on blockchain training. It is a great step, and we hope to see more colleges and institutions join to increase hands-on training.

\subsection{Blockchain market growth}
The blockchain market is expected to continue growing in the coming years, driven by increasing demand for blockchain-based solutions and the growing adoption of blockchain technology by businesses and organizations worldwide \cite{R44}. According to some estimates \cite{N10}, the global blockchain market is projected to reach a value of over \$39 billion by 2025. In addition, it is growing at a compound annual growth rate of 67.3\% between 2020 and 2025. The blockchain market is highly competitive, with many companies and organizations offering various blockchain-based products and services. Some of the key players in the market include IBM, Microsoft, Oracle, AWS, and Accenture. 

Global Fintech blockchain market size was valued at \$ 15.51 billion in 2021 and is projected to reach \$ 37.44 billion by 2028, growing at a CAGR of 50.3\% from 2021 to 2028 according to a new report by Intellectual Market Insights Research \cite{N2}. The Fintech blockchain market research report examines this industry's existing and future state in depth \cite{N3}. While broadening the company's view, the research identifies important trends, possibilities, and obstacles for numerous segments and sub-segments. The study report also contains substantial information based on historical and present patterns across different industry verticals to aid in the identification of possible expansion opportunities. The forecast period offers several market share, size, and industry growth rate estimates. According to the Fintech blockchain industry \cite{N2}, the study contains information on competitor analysis, consumption habits, and price tactics.



Further, we take a look at the following predictions of how blockchain technology will influence various sectors of the global landscape \cite{R18}.

{\raggedright \textbf{a. Governance will use blockchain application}}

One trend in blockchain technology is its application in government. Governments must prepare for the eventual takeover of bitcoin as markets get saturated. Governments can also gain from the use of virtual currency. However, each government department has its have database, making information about other databases difficult. Furthermore, governments are usually required to comply with rules. This makes compliance challenging for multinational corporations \cite{R58}.

On the other hand, future blockchain technology advancements will improve the functionality of effective government data management. As a result, these agencies can manage large volumes of data more efficiently.

{\raggedright \textbf{b. NFT expanding beyond online art}}

Non-Fungible Tokens (NFTs) were the hottest topic in the blockchain industry in 2021 \cite{R2}. Artworks like Beeple's first 5000 days fetched exorbitant prices, solidifying the public's understanding of unique digital tokens stored on blockchains. NFTs are also popular and play a vital in the music industry, with artists like Kings of Leon, Shawn Mendes, and Grimes all releasing NFT-formatted compositions. However, like the blockchain technology itself, the concept has promise beyond its first attention-getting applications.

{\raggedright \textbf{c. Blockchain will increase in retail}}

Blockchain innovation trends in 2022 will influence the inventory network business due to increased emphasis on exchanges \cite{N4}. Supply chains have recently gotten extraordinarily complex \cite{R37}. Because of the chaos of sending and receiving installments, specialists must make many movements to discover criminals. Furthermore, stock management between the provider, retailer, and the manufacturer might mask illegal activity. The mediators on the store network board provide a significant risk to the specialists and retailers in charge of trader administrations and Visa processing.

The retail industry is expected to grow significantly; thus, robust stock management and store network systems must be in place. Furthermore, many believe blockchain will rescue retail long after the epidemic has passed. As the demand for traceable, flexible client help and operations grows, the retail industry will be consumed by blockchain innovation.

{\raggedright \textbf{d. Blockchain and IoT integration}}

Because it is ideal for maintaining records of interactions and transactions between devices, blockchain is a wonderful fit for the Internet of Things (IoT) \cite{N5}. Blockchain ledgers and databases have the potential to solve several security and scalability challenges because of their automated, encrypted, and irreversible nature. It might also be used for machine-to-machine transactions, allowing cryptocurrencies to be used to make micropayments when one machine or network wants to acquire services from another. While this is an advanced use case that may require us to travel farther down the road before it influences our daily lives, we will hear about additional pilot projects and early use cases in this subject in 2022. Furthermore, the implementation of 5G networks is anticipated to stimulate innovation in this space, implying more connectedness between all kinds of smart, networked devices and appliances — not just in terms of speed but also in terms of new types of data, such as blockchain transactions \cite{R56}.

{\raggedright \textbf{e. Big Tech will enhance using BaaS}}

Big Tech companies may use blockchain technology in their operations. Furthermore, Microsoft and Amazon have already invested in and are promoting emergency blockchain technology as a service. BaaS, or blockchain-as-a-service, is a cloud-based service that allows customers to create their blockchain-based digital products \cite{N6}.

These items are often smart contracts or dapps operating in the blockchain architecture without setup requirements removing the heavy-lifting configuration of blockchain networks or nodes.

{\raggedright \textbf{f. Social media will rely on blockchain}}

As if social media was not already widespread enough, blockchain will soon join the party. User identity verification and marketplace verification were two important social media-related blockchain technology topics in 2022. For example, bots are a well-known problem on social media sites. This is not only a political problem but also a commercial one. Marketers squander time on inactive profiles, necessitating the usage of blockchain technology and smart contracts to verify user identity \cite{N7}.

Furthermore, marketplace verification, another upcoming development in blockchain technology, will boost a company's growth potential. This allows consumers to promote to certified merchants, simplifying the marketing process. Finally, incorporating blockchain into social media will aid in verifying publicly available data. As a result, it is untraceable and cannot be copied even after it has been created \cite{R59}.

{\raggedright \textbf{A summary of HEIs that have the best blockchain programs}}

Many colleges provide education on blockchain technology to students because the demands in the industry are increasing rapidly \cite{R9}. It is the technology that has wider scope in the present and also in the future. In today's scenario, blockchain is a revolutionary technology.

Colleges teach blockchain technology to students because it helps them learn about different concepts of blockchain. Many private institutions offer full-time and part-time programs, courses, and certifications on various blockchain technologies \cite{R46}.

For example, undergraduates at Berkeley can study blockchain fundamentals and blockchain software development through the industrial engineering and computer science departments, respectively.

The blockchain at Berkeley group on campus orchestrates courses for students to increase familiarity with the technology and provide the tools needed to develop applications. Other courses look at substantiating blockchain use cases \cite{R49}. The group also provides external consulting services to existing blockchain firms.

This list of courses will get bigger very soon, with most universities with strong computing departments striving to provide courses in the blockchain \cite{R7}. However, a few leading universities that already offer these courses are the following:
\begin{enumerate}
\item Illinois (Cryptocurrency security)
\item MIT (Shared public ledgers, cryptocurrencies)
\item George Mason (Blockchain Technologies)
\item Stanford (Bitcoin Engineering)
\item  Duke (Innovation and Cryptoventures)
\end{enumerate}

Indeed, quite a few universities are trialing courses and providing them online. For instance, Princeton university students will prefer to work in a more university-oriented environment but not have to sit through the whole module length. Other universities, like the University of Copenhagen, offer a Blockchain Summer School, which cooperates with the European Blockchain Center, featuring workshops and seminars to enhance students' understanding of blockchain technology \cite{R47}.


\section{Future Directions}
\label{sec:FD}
Blockchain technology is so vast that it keeps updating with new techniques and trends. So HEIs have to keep students updated now and then. Many blockchain experts in this industry keep themselves tuned with the latest blockchain technology standards. Even though there is a plethora of information available in the Public domain about blockchain technology, for a beginner, there is no easy way to segregate this information and ramp up his/her learning.

Since this technology has already branched out into numerous industries, it is expected to expand even more. These industries, and the businesses that operate inside them, are on the lookout for professional developers who can use and deploy blockchain technology. The future of a blockchain developer is bright, and it is unlikely to cover shortly.

\textit{"Banking, security, real estate, education, healthcare, supply chain, and now even voting have begun and will continue to use blockchain technology in their operations. The scope for the professional developer, architect, engineer will open up many options in many areas and businesses. The developer and architects will have a big number of prospects"} \cite{N14}

The developer can innovate and produce even more remarkable applications to benefit the global economy and industries. 




When talking about the future scope of the blockchain profession, the interesting fact is that according to the latest skills index, blockchain expertise is the fastest-growing skill and is now one of the hottest in the United States Job market \cite{R21}. This craze of blockchain will never stop; if students want to pursue a career in blockchain, it is the right time!

Many blockchain courses, including solution architects, designers, and engineer, have a precarious and somewhat checkered history in academia. Curricula in universities need to include blockchain courses largely, as mentioned earlier. Instead, the teachings are often left to the students to proceed with external sources to study. Over time this needs to change by the universities as the blockchain market grows, attracting attention from top talent in engineering and business. Blockchain education absence from university is not a problem with the blockchain workforce itself but rather with its insufficient embrace of innovation. 
Blockchain can serve as an inspiration for what academic research can and should be. In fact, it presents a roadmap to improve higher education.





\section{Conclusion}
\label{sec:con}
This paper discussed the current state of blockchain education and the problems it faces in meeting industry workforce demand. In order to respond to the growing demand for blockchain professionals, we investigated various career tracks available for blockchain learners and the necessary certifications and courses to educate blockchain technology. In short, we listed various blockchain applications' trends, their importance in industries, and changes needed in HEIs. In addition, we discussed how blockchain education should be more comprehensive than computing since this technology capitalizes on many different discipline applications. The future suggests developing curricula based on required roles-related skills (in terms of course or program). 
\bibliographystyle{elsarticle-num}
\bibliography{mybibfile}
\end{document}